\documentclass[apj]{emulateapj}
\usepackage{apjfonts}

\shorttitle{DM EQUILIBRIA IN GALAXIES AND GALAXY SYSTEMS}
\shortauthors{LAPI \& CAVALIERE}
\journalinfo{Accepted by ApJ}

\begin{document}
\title{Dark Matter Equilibria in Galaxies and Galaxy Systems}
\author{A. Lapi\altaffilmark{1,2} and A. Cavaliere\altaffilmark{1,3}}
\altaffiltext{1}{Univ. `Tor Vergata', Via Ricerca Scientifica
1, 00133 Roma, Italy.} \altaffiltext{2}{Astrophysics Sector,
SISSA/ISAS, Via Beirut 2-4, 34014 Trieste, Italy.}
\altaffiltext{3}{Accademia Nazionale dei Lincei, Via Lungara
10, 00165 Roma, Italy}

\begin{abstract}
In the dark matter (DM) halos embedding galaxies and galaxy
systems the `entropy' $K\equiv \sigma^2/\rho^{2/3}$ (a quantity
that combines the radial velocity dispersion $\sigma$ with the
density $\rho$) is found from intensive $N-$body simulations to
follow a powerlaw run $K \propto r^{\alpha}$ throughout the
halos' bulk, with $\alpha$ around $1.25$. Taking up from
phenomenology just that $\alpha\approx $ const applies, we cut
through the rich analytic contents of the Jeans equation
describing the self-gravitating equilibria of the DM; we
specifically focus on computing and discussing a set of novel
physical solutions that we name $\alpha$-\emph{profiles},
marked by the entropy slope $\alpha$ itself, and by the maximal
gravitational pull $\kappa_{\mathrm{crit}}(\alpha)$ required
for a viable equilibrium to hold. We then use an advanced
semianalytic description for the cosmological buildup of halos
to constrain the values of $\alpha$ to within the narrow range
$1.25-1.29$ from galaxies to galaxy systems; these correspond
to halos' current masses in the range $10^{11}-10^{15}\,
M_{\odot}$. Our range of $\alpha$ applies since the transition
time that $-$ both in our semianalytic description and in
state-of-the-art numerical simulations $-$ separates two
development stages: an early violent collapse that comprises a
few major mergers and enforces dynamical mixing, followed by
smoother mass addition through slow accretion. In our range of
$\alpha$ we provide a close fit for the relation
$\kappa_{\mathrm{crit}}(\alpha)$, and discuss a related
physical interpretation in terms of incomplete randomization of
the infall kinetic energy through dynamical mixing. We also
give an accurate analytic representation of the
$\alpha$-profiles with parameters derived from the Jeans
equation; this provides straightforward \emph{precision} fits
to recent detailed data from gravitational lensing in and
around massive galaxy clusters, and thus replaces the empirical
NFW formula relieving the related problems of high
concentration and old age. We finally stress how our findings
and predictions as to $\alpha$ and $\kappa_{\mathrm{crit}}$
contribute to understand hitherto unsolved issues concerning
the fundamental structure of DM halos.
\end{abstract}
\keywords{Dark matter -- galaxies: clusters: general -- galaxies: halos}

\section{Introduction}
Galaxies, galaxy groups and clusters are widely held to form
under the drive of the gravitational instability that acts on
initial perturbations modulating the cosmic density of the
dominant cold dark matter (DM) component. The instability at
first is kept in check by the cosmic expansion, but when the
local gravity prevails collapse sets in. The standard sequence
runs as follows: a slightly overdense region expands more
slowly than its surroundings, progressively detaches from the
Hubble expansion, halts and turns around; then it collapses,
and eventually virializes to form a DM `halo' in equilibrium
under self-gravity. The amplitude of more massive perturbations
is smaller, so the formation is hierarchical with the massive
structures forming typically later (see Peebles 1993).

Such a formation history has been confirmed and resolved to a
considerable detail by many intensive $N-$body simulations.
Early on (White 1986) these added an important block of
information, i.e., the growth of a halo actually includes
merging events with other clumps of sizes comparable (`major'
mergers) or substantially smaller (`minor' mergers), down to
nearly smooth accretion (see Springel et al. 2006).

More recently, a second round of features has been increasingly
recognized in highly resolved simulations of individual halos
forming in cosmological volumes (Zhao et al. 2003; Diemand et
al. 2007), to the effect of identifying in the growth two
definite stages: an early fast collapse including a few violent
major mergers, and a later calmer stage comprising many minor
mergers and smooth accretion. During the early collapse a
substantial mass is gathered, while the major mergers reshuffle
the gravitational potential wells and cause the collisionless
DM particles to undergo dynamical relaxation; therefrom the
halo emerges with a definite structure of the inner density and
gravitational potential. During the later stage, moderate
amounts of mass are slowly accreted mainly onto the outskirts,
little affecting the inner structure and potential, but
quiescently rescaling upwards the overall size (see Hoffman et
al. 2007).

In the resulting simulated halos, the coarse-grained proxy
$Q\equiv\rho/\sigma^3$ to the phase-space density -- defined in
terms of the DM density $\rho$ and of the radial velocity
dispersion $\sigma$ -- is empirically found to be stable after
the collapse, and distributed in the form of a powerlaw with
\emph{uniform} slope $\mathrm{d}\log Q/\mathrm{d}\log r$ around
$-1.9$ throughout the halo's bulk where it is amenable to
noise-free measurements (see Taylor \& Navarro 2001; Dehnen \&
McLaughlin 2005; Ascasibar \& Gottl\"{o}ber 2008).

Taking up such a notion, the present paper starts (\S~2) from
investigating the wide set of macroscopic equilibria for the DM
under self-gravity, as described by the Jeans equation. It
proceeds (\S~3) to the semianalytic description of their time
buildup to constrain their governing parameters, and so predict
the halo realistic structure. It proposes (\S~4) physical
interpretations for such parameters. The paper is concluded
(\S~5) with a summary of our findings and a discussion of the
perspectives they open.

Throughout we adopt the standard, flat cosmological model with
normalized matter density $\Omega_M = 0.27$, dark energy
density $\Omega_\Lambda = 0.73$, and Hubble constant $H_0 = 72$
km s$^{-1}$ Mpc$^{-1}$.

\begin{figure*}
\epsscale{0.8}\plotone{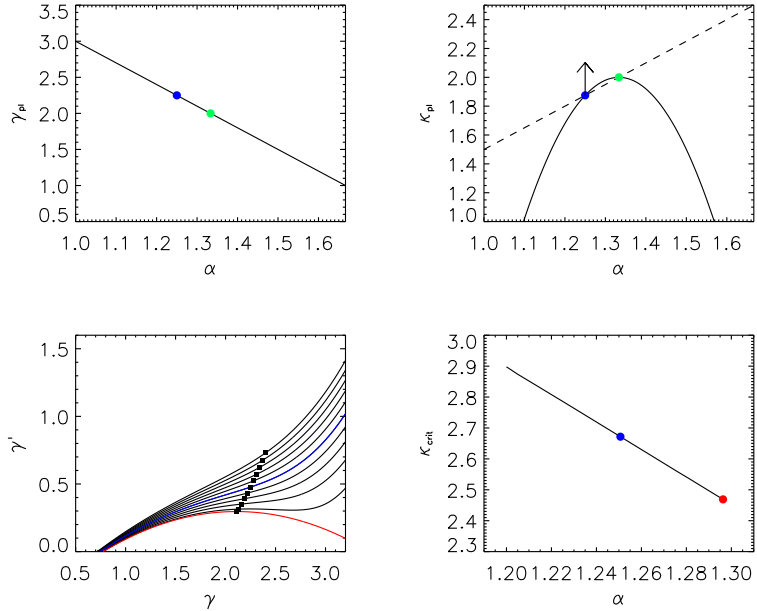}\caption{Top panels: the
$\gamma(\alpha)$ and $\kappa(\alpha)$ relations for the
powerlaw solutions of the Jeans equation, see \S~2.1; blue dots
refer to the solution with $\alpha=5/4$, and green dots refer
to the solution with $\alpha=4/3$ (the isothermal sphere). In
the right panel the \textit{dashed} line represents the
relation $\kappa=3\, \alpha/2$, and the arrow indicates how
$\kappa$ increase toward the values (given in the bottom right
panel) for the $\alpha$-profiles, see \S~2.1-2.2 for details.
Bottom panels: the $\gamma-\gamma'$ and
$\kappa_{\mathrm{crit}}(\alpha)$ relations for the
$\alpha$-profiles with $\alpha$ in the range $1.2-35/27$, see
\S~2.1-2.2; blue dots and curves refer to the solution with
$\alpha=5/4$, and red dots and curves refer to the solution
with $\alpha=35/27$. In the left panel, the squares mark the
couple of values $\gamma_0$, $\gamma_0'$ at $r=r_0$. Note that
we will find in \S~3 the relevant values of $\alpha$ to be
constrained within the range $5/4-35/27$, i.e., between the
blue and red dots/curves.}
\end{figure*}

\begin{figure*}
\epsscale{0.8}\plotone{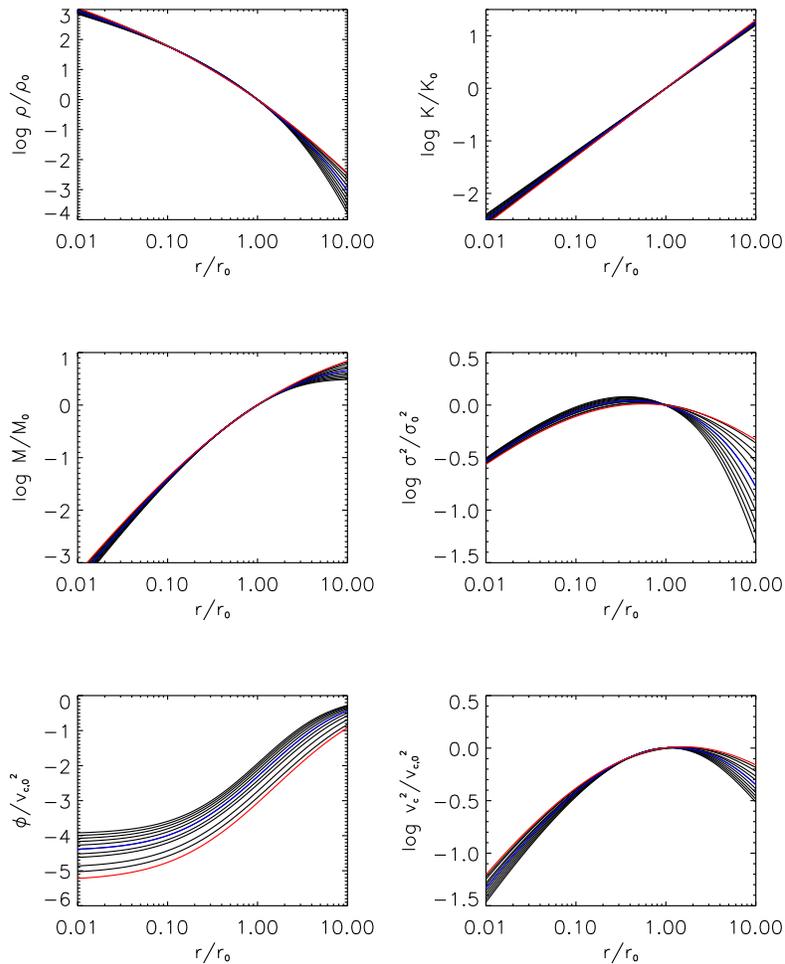}\caption{Radial runs for
the $\alpha$-profiles: density $\rho$, mass $M$, circular
velocity $v_c$, gravitational potential $\Phi$, velocity
dispersion $\sigma$, and entropy $K$; the profiles are
normalized to $1$ at the point $r_0$ where
$\gamma=\gamma_0=6-3\,\alpha$ holds. Various curves are for
$\alpha$ in the range $1.2-35/27$; the blue one refers to the
solution with $\alpha=5/4$, and the red one refers to
$\alpha=35/27$.}
\end{figure*}

\begin{figure*}
\epsscale{0.8}\plotone{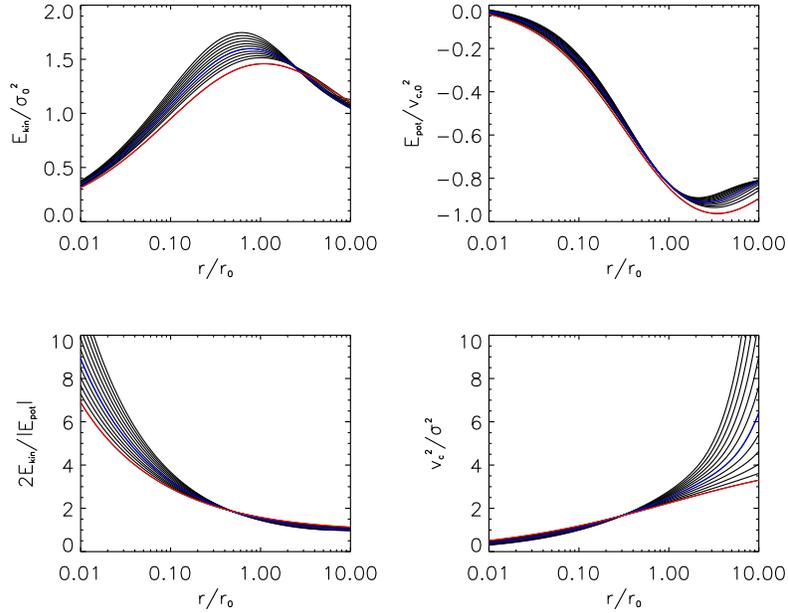}\caption{Energy
distributions for the $\alpha$-profiles: specific potential
energy $E_{\mathrm{pot}}$, random kinetic energy
$E_{\mathrm{kin}}$, virial ratio
$2\,E_{\mathrm{kin}}/|E_{\mathrm{pot}}|$, and ratio of circular
velocity to the velocity dispersion $v_c^2/\sigma^2$ appearing
in Eq.~(2). Lines as in Fig.~2.}
\end{figure*}

\section{Jeans equilibria}

The static equilibria of the DM halos obey the Jeans equation
that, on neglecting for the time being anisotropies and the
blend into the large-scale environment to be discussed in \S~5,
reads
\begin{equation}
{1\over \rho} {\mathrm{d} (\rho\, \sigma^2)\over
\mathrm{d}r}= -\, {GM(<r)\over r^2}~.
\end{equation}

To elicit the contents of this macroscopic-like equation, an
explicit link is required between $\sigma^2(r)$ and $\rho(r)$.
Such a link may be provided by $Q(r)$, or by the equivalent
quantity $K(r)\equiv \sigma^2/\rho^{2/3}=Q^{-2/3}(r)$; this is
often named DM `entropy' in view of its formal analogy with the
adiabat of a gas in thermal equilibrium, and is increasingly
used in the context of DM equilibria, from Taylor \& Navarro
(2001) to Faltenbacher et al. (2007). We focus on this function
of state $K(r)$ in the configuration space, and comment on our
choice in \S~5.

In terms of the slope $\mathrm{d}\log K/\mathrm{d}\log r\equiv
\alpha$, Eq.~(1) is recast into the particularly simple form
\begin{equation}
{5\over 3}\, {\mathrm{d}\log\, \rho\over \mathrm{d}\log\, r} =
- {\mathrm{d} \log \,K \over \mathrm{d}\log \, r}  -
{v^2_c(r)\over \sigma^2(r)} ~.
\end{equation}
Here $v_c^2 (r)= G\, M(<r)/r$ is the (squared) circular
velocity, that provides a natural scaling for the depth of the
DM potential well.

Taking up from the simulation phenomenology only the notion
that $\alpha$ be \emph{uniform} throughout the halo bulk (see
\S~1), we will discuss the range of values to be expected for
$\alpha$ as a clue to their origin. In fact, in their
pioneering work Taylor \& Navarro (2001) selected a value of
the exponent close to $\alpha = 1.25$. Other, recent studies
have provided similar values for $\alpha$: around $1.27$
according to Ascasibar et al. (2004), and around $1.29$
according to Rasia et al. (2004). Beyond the minor differences,
it is the \emph{narrow} empirical range that constituted an
unsolved challenge.

\subsection{Background}

We begin with looking first (after Taylor \& Navarro 2001) at
powerlaw solutions of Eq.~(2) of the form $\rho\propto
r^{-\gamma}$, with the entropy run set to $K\propto
r^{\alpha}$; then not only $\alpha$ is a given parameter, but
also the slope $\gamma = -\mathrm{d} \log \rho/\mathrm{d}\log
r$ is a \emph{constant} to be determined from Eq.~(2) in the
form
\begin{equation}
\gamma -{3\over 5}\, \alpha =  {3\over 5}\,{\kappa\over
3-\gamma}\, \left({r\over r_0}\right)^{2-\alpha - \gamma/3 }~.
\end{equation}
Here the `bare' constant $\kappa\equiv
4\pi\,G\rho_0\,r_0^2/\sigma_0^2=(3-\gamma)\,v_0^2/\sigma_0^2$
compares the gravitational potential to the random kinetic
energy associated with the velocity dispersion at a reference
point $r_0$. At the r.h.s. $\kappa$ appears as modulated by
$r-$dependencies, produced by the integrated dynamical mass
$M(<r)$ to yield $v^2_c (r) \propto r^{2 - \gamma}/(3-\gamma)$
in the powerlaw case, and by the local dependencies from
$\sigma^2(r) \propto K(r) \, \rho^{2/3}(r) \propto r^{\alpha -
2\gamma/3}$.

For a powerlaw solution to actually hold the r.h.s. must be
constant, which yields the link
\begin{equation}
\gamma=6-3\,\alpha~,
\end{equation}
as pointed out by Taylor \& Navarro (2001), Hansen (2004), and
Austin et al. (2005). Given this, the balance of the three
terms in Eq.~(3) imposes on the dimensionless gravitational
pull the constraint
\begin{equation}
\kappa = (5\gamma/3-\alpha)(3-\gamma) = 6\,(\alpha-1)\,(5-3\,\alpha)~,
\end{equation}
whose nonlinearities reflect the DM self-gravity. Thus the
allowed powerlaw solutions must lie on a parabola in the plane
$\alpha-\kappa$ and on a straight line in the plane
$\alpha-\gamma$, as illustrated in the top panels of Fig.~1;
note that to ensure $\kappa>0$ the overall bounds
$1<\alpha<5/3$ must hold, implying that $1<\gamma<3$ applies
for the powerlaw solutions.

Two notable instances correspond to the values $\alpha=4/3$ and
$5/4$. The former yields the standard, singular isothermal
sphere with $\gamma=2$ and $\kappa=2$; the latter holds for the
value of $\alpha$ originally selected by Taylor \& Navarro
(2001), and implies $\gamma=9/4$ and $\kappa=15/8$. Note that
they are located on the intersections of the parabola Eq.~(5)
with the line $\kappa=3\,\alpha/2$, see Fig.~1.

The powerlaw solutions of the Jeans equation with slope
$\gamma_0\equiv 6-3\,\alpha$ will turn out to be tangent to the
full solutions at $r = r_0$ so as to provide a useful reference
slope in the middle of the halo. However, they do not qualify
as relevant solutions since at both extremes they suffer of
unphysical features: since $1<\gamma<3$ holds as shown above,
the force $G\,M(<r)/r^2\propto r^{1-\gamma}$ would diverge at
the center, while the mass $M(<r)\propto r^{3-\gamma}$ would
diverge toward large $r$.

To avoid these drawbacks, in all physically relevant solutions
the running slope $\gamma(r)$ is to bend down from the middle
value in going toward large and small $r$. In fact, for $r\ll
r_0$ one can infer from Eq.~(3) that asymptotically
$\gamma\rightarrow \gamma_a\equiv 3\, \alpha/5$ is to hold if
the gravitational force is to be finite, which implies that
both sides of Eq.~(3) vanish (see Austin et al. 2005); on the
other hand, for $r\gg r_0$ it is seen that either
$\gamma\rightarrow \gamma_b\equiv 3\,(1+ \alpha)/2$ holds or a
density cutoff occurs, if the mass $M(<r)$ is to saturate to a
constant, implying $v^2_c (r)\propto 1/r$. In other words,
powerlaws with inner $\gamma_a$ and outer slope $\gamma_b$ will
constitute asymptotic or effective approximations to the full
solutions. For example, for the value $\alpha= 5/4$ one
recovers the slopes $\gamma_a= 3/4$ and $\gamma_b=27/8$, found
by Taylor \& Navarro (2001) in their archetypical full
solution.

Thus all full solutions may be conveniently visualized as
\emph{bent down} from the corresponding tangent powerlaw so as
to meet the above asymptotic forms at both small and large
radii. These solutions will be analytically studied and
constrained next. In fact, the solution space of the Jeans
equation is analytically rich, so we refer the interested
reader to the works by Austin et al. (2005), Dehnen \&
McLaughlin (2005), and Barnes et al. (2006, 2007) for extensive
coverage. Here we summarize the main results and proceed to
focus on the `critical' solutions in the terminology of Taylor
\& Navarro (2001).

To highlight the properties of these solutions with running
$\gamma(r)$, it is convenient (see Dehnen \& McLaughlin 2005;
Austin et al. 2005; also Barnes et al. 2006, 2007) to replace
the original Jeans equation $\gamma-3\,\alpha/5 = 3\,v_c^2/5\,
\sigma^2$ with its first and second derivative, and use the
differential mass definition
$\mathrm{d}M/\mathrm{d}r=4\pi\rho\,r^2$, to obtain
\begin{eqnarray}
&\nonumber &\gamma'-{2\over
3}\,(\gamma-\gamma_a)\,(\gamma-\gamma_b)={3\over
5}\,\kappa\,\left({r\over
r_0}\right)^{2-\alpha}\,\left({\rho\over
\rho_0}\right)^{1/3}~,\\
& &\\
&\nonumber &\gamma''-\gamma'\,\left(\gamma-{2\gamma_a+2\,\gamma_b-\gamma_0\over
3}\right)={2\over
9}\,(\gamma-\gamma_a)\,(\gamma-\gamma_0)\,(\gamma-\gamma_b)~.
\end{eqnarray}
Here $\gamma'$ and $\gamma''$ denote the first and second
logarithmic derivative of $\gamma$ with respect to $r$; in
addition, $r_0$ is the point where the density slope matches
$\gamma_0\equiv 6-3\,\alpha$ (see Eq.~[4]), while
$\gamma_a\equiv 3\,\alpha/5$ and $\gamma_b\equiv
3\,(1+\alpha)/2$ are shorthands for the inner and the outer
effective slopes anticipated above.

In the second equation the normalization constant $\kappa\equiv
4\pi G\rho_0 r_0^2/\sigma_0^2$ does not appear explicitly;
however, from the first Eq. (6) evaluated at $r_0$ it is easily
recognized that
\begin{equation}
\kappa=6\,(\alpha-1)\,(5-3\,\alpha)+5\,\gamma'_0/3
\end{equation}
is related to $\gamma'_0$, and in a full solution specifically
measures the upward deviation of $\kappa$ from the
corresponding powerlaw, that is, above the parabola expressed
by Eq.~(5), see also Fig.~1.

In fact, as the value of $\kappa$ increases at given $\alpha$,
the solutions will be progressively curved down both inward and
outward of $r_0$, to constitute a bundle \emph{tangent} to the
related powerlaw with slope given by Eq.~(4). But, as stressed
by Taylor \& Navarro (2001), only a definite maximal value
$\kappa_{\mathrm{crit}}$ can be withstood by the halo at
equilibrium, lest the solution develops pathologies like a
central hole; this value marks a `critical' solution.
Correspondingly, the values for
$\kappa_{\mathrm{crit}}(\alpha)$ may be found by a painstaking
trial-and-error procedure, as done by Taylor \& Navarro (2001)
for the particular case $\alpha=5/4$. More conveniently, they
may be computed directly through Eqs.~(6) and (7); the former
also provides the following exhaustive description of the
solutions.

In the \emph{inner} region only three situations possibly arise
(Williams et al. 2004, Dehnen \& McLaughlin 2005): (i) $\gamma$
attains the value $\gamma_0$ asymptotically for $r\ll r_0$,
which is unphysical since such a  solution would end there with
a diverging gravitational force; (ii) the solution attains the
value $\gamma_a$ at a finite radius, and then continues to
flatten down to the point of developing a central hole, which
is obviously to be avoided; (iii) $\gamma$ approaches the value
$\gamma_a$ asymptotically for $r\ll r_0$, that is the only
\emph{viable} possibility. Thus physical solutions must
approach the inner slope $\gamma_a$, and must steepen
monotonically outwards.

However, not all of them are tenable in view of their
\emph{outer} behavior (Austin et al. 2005, Dehnen \& McLaughlin
2005). Here, again three situations formally arise that depend
on the value of $\alpha$: (i) $\gamma$ approaches the value
$\gamma_0$ asymptotically for $r\gg r_0$, which occurs for
$\alpha>35/27$ and is unphysical since with such a solution the
mass would diverge (in practice, it would be ill-defined, see
\S~5); (ii) $\gamma$ attains the slope $\gamma_b$ at a finite
radius and then goes to an outer cutoff, that occurs for
$\alpha<35/27$ and is physically \emph{viable} since the
cutoffs actually set in at large radii exterior to the halo's
bulk, as discussed below; (iii) $\gamma$ approaches the slope
$\gamma_b$ asymptotically for $r\rightarrow \infty$ as it
occurs for $\alpha=35/27$, which also provides a physically
\emph{viable} solution.

\begin{figure*}
\epsscale{0.8}\plotone{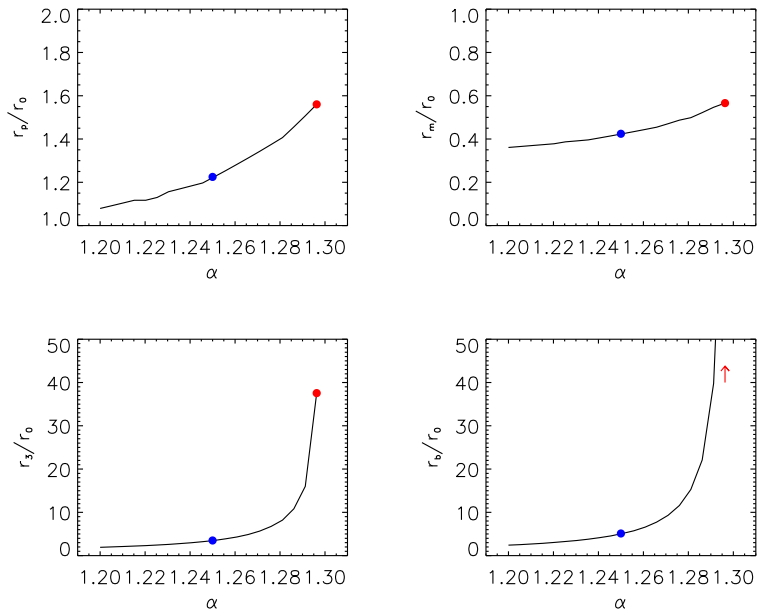}\caption{Characteristic
radii (in units of $r_0$) for the $\alpha$-profiles as a
function of $\alpha$: radius $r_p$ where the circular velocity
peaks, radius $r_m$ where the velocity dispersion peaks, radius
$r_3$ where the slope of the density profile equals $3$, and
radius $r_{b}$ where the slope of the density profile reads
$\gamma_b=3\,(1+\alpha)/2$. The blue dots refer to the solution
with $\alpha=5/4$, and the red dots to $\alpha=35/27$. In the
bottom right panel the red arrow indicates that the solution
with $\alpha=35/27$ attains the slope $\gamma_b$ only
asymptotically in a powerlaw tail. Note that as $\alpha$
increases, the peak of the velocity dispersion displaces
relatively more than the circular velocity's.}
\end{figure*}

\subsection{Viable full solutions}

Summing up, the critical solutions satisfy both sets of
criteria, but exist only for $\alpha\le 35/27 =
1.\overline{296}$ (which excludes the case corresponding to the
isothermal powerlaw); hereafter these will be named
`$\alpha$-\emph{profiles}'. We compute them by integrating the
second of Eqs.~(6) with boundary conditions $\gamma\rightarrow
\gamma_a$ and $\gamma'\rightarrow 0$ for $r\rightarrow 0$. The
results in the plane $\gamma-\gamma'$ for various $\alpha$
between $1.2$ and $35/27$ are plotted as different lines in
Fig.~1 (left bottom panel); there we can read out the values of
$\gamma'_0$ at $r=r_0$ (squares in the figure), and find the
related $\kappa_{\mathrm{crit}}$ after Eq.~(7). The relation
$\kappa_{\mathrm{crit}}(\alpha)$ is plotted in Fig.~1 (right
bottom panel); the resulting correlation is fitted by the
expression $\kappa_{\mathrm{crit}}(\alpha)\simeq
8.228-4.443\,\alpha$ to better than $1\%$.

Our solutions include the prototypical one corresponding to
$\alpha=5/4$ (highlighted in blue), first found by Taylor \&
Navarro (2001) and shown to be well fitted, if only in a region
around $r_0$, by the NFW formula (Navarro et al. 1997); for
such a solution the precise value of $\kappa_{\mathrm{crit}}$
reads $2.674$, see Dehnen \& McLaughlin (2005). Our solutions
also include the one corresponding to the limiting value
$\alpha=35/27$ (highlighted in red) studied by Dehnen \&
McLaughlin (2005); for this $\kappa_{\mathrm{crit}}=200/81$ is
found.

In Fig.~2 we illustrate the radial dependencies of various
interesting quantities for the $\alpha$-profiles: density
$\rho$ and its logarithmic slope $\gamma$, circular velocity
$v^2_c$, velocity dispersion $\sigma^2$, entropy $K$, mass
$M(<r)$, and gravitational potential
$\Phi=-\int_r^{\infty}{\mathrm{d}r'}\,G\,M(<r')/r'^2$; the
profiles are normalized to $1$ at the point $r_0$ within the
halo's middle. Outwards of this, solutions corresponding to
larger values of $\alpha$ have larger overall masses, flatter
density profiles, higher velocity dispersions, and deeper
potential wells.

As mentioned above, the $\alpha$-profiles with $\alpha < 35/27$
show an outer cutoff, to which the following remarks apply.
First, the cutoffs generally occur beyond the virial radius
(see \S~5). Second, the velocity dispersion $\sigma^2 \propto
\rho^{2/3} \, K\propto r^{\alpha-2\gamma/3}$ falls rapidly
outside of $r_0$ and vanishes at the density cutoff, thus
preventing any finite outflow related to residual random
motions of DM particles. Third, the microscopic distribution
function as obtained through the Eddington formula (e.g.,
Binney \& Tremaine 1987) from pairs of $\rho(r)$ and $\Phi(r)$
is realistic being nowhere negative (see also Zhao 1996; Widrow
2000).

In Fig.~3 we show the ratio $v_c^2(r)/\sigma^2(r)$ entering
Eq.~(2); we also show the energy distributions associated to
the $\alpha$-profiles: the specific potential energy
$E_{\mathrm{pot}}=4\pi\int_0^r{\mathrm{d}r'}\,$
$r'^2\,\rho(r')\,v_c^2(r')/M(<r)$, the random kinetic energy
$E_{\mathrm{kin}}=4\pi\int_0^r{\mathrm{d}r'}\,r'^2\,\rho(r')\;
3\,\sigma^2(r')/2\,M(<r)$, and the local virial ratio
$2\,E_{\mathrm{kin}}/|E_{\mathrm{pot}}|$. Note that the latter
tends to $1$ at large $r$, as is expected for stable
self-gravitating halos (see {\L}okas \& Mamon 2001); in fact,
it closely approaches unity already for $r\approx r_0$.

In Fig.~4 we plot as a function of $\alpha$ various
characteristic radii (in units of $r_0$) of the
$\alpha$-profiles; specifically, we show the radius $r_p$ where
the circular velocity $v_c(r)$ peaks, the radius $r_m$ where
the velocity dispersion $\sigma^2(r)$ peaks, the radius $r_3$
where the slope of the density profile reads $\gamma=3$, and
the radius $r_{b}$ where the slope of the profile takes on the
value $\gamma_b=3\,(1+\alpha)/2$. Note that $r_b$ is always
large, and formally recedes to infinity (as represented in the
plot with an arrow) for the limiting case $\alpha=35/27$ that
has a powerlaw falloff.

For the sake of clarity, in these Figures we have confined the
values of $\alpha$ to between $1.2$ and $35/27$; although the
upper bound is already required for having no central hole and
a finite mass as discussed above, limited information existed
so far concerning lower bounds from the isotropic Jeans
equation.

\section{The entropy slopes from \\cosmogonic buildup}

Here we show that the cosmogonic \emph{histories} of DM halos
enforce additional, stringent \emph{limitations} to the range
of $\alpha$; on the high side the resulting bound turns out to
be pleasingly close to the value $35/27$ derived above from the
analysis of the static Jeans equation, but on the low side it
constitutes a new stringent result.

Consider a DM halo with current bounding radius $R$, density
$\rho$, mass $M\propto \rho\,R^3$, velocity scales
$\sigma^2\propto v_c^2\propto M/R$, and entropy $K\propto
\sigma^2/\rho^{2/3}$. During the hierarchical buildup the
running growth rate scales as $\dot{M}\propto \rho\,v\,R^2$
with the inflow speed $v$ set by the depth of the DM well and
hence proportional to $v_c^2$; thus the above quantities may be
expressed in terms of $M$ and $\dot{M}$ to provide the scaling
laws:
\begin{equation}
R\propto M/\dot{M}^{2/3}~~~~~~v_c^2\propto \sigma^2 \propto
\dot{M}^{2/3}~~~~~~\rho\propto
\dot{M}^2/M^2~~~~~~K\propto
M^{4/3}/\dot{M}^{2/3}~.
\end{equation}

\begin{figure*}
\epsscale{0.8}\plotone{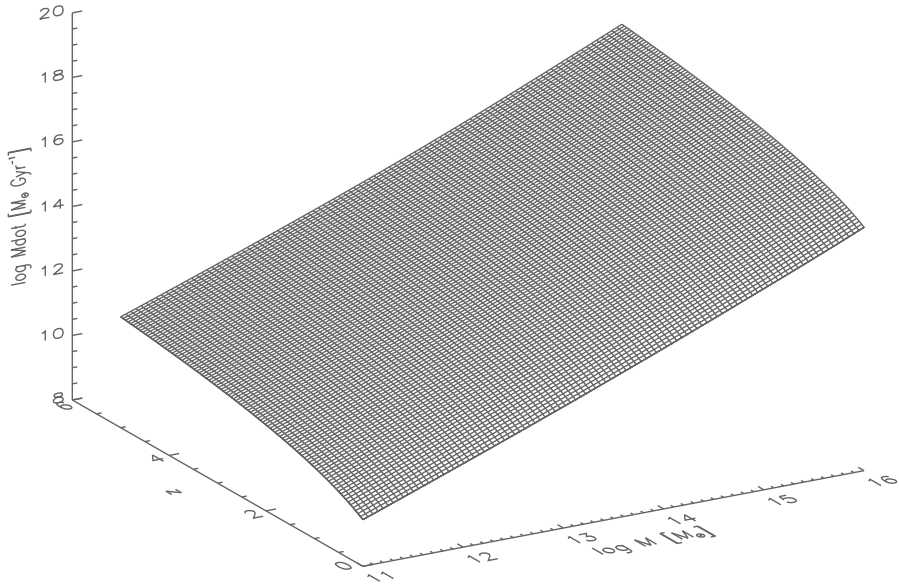}\caption{The relation
$\dot{M}(M,z)$ from the average cosmogonic evolution, see
\S~3.2 for details.}
\end{figure*}

\begin{figure*}
\epsscale{0.8}\plotone{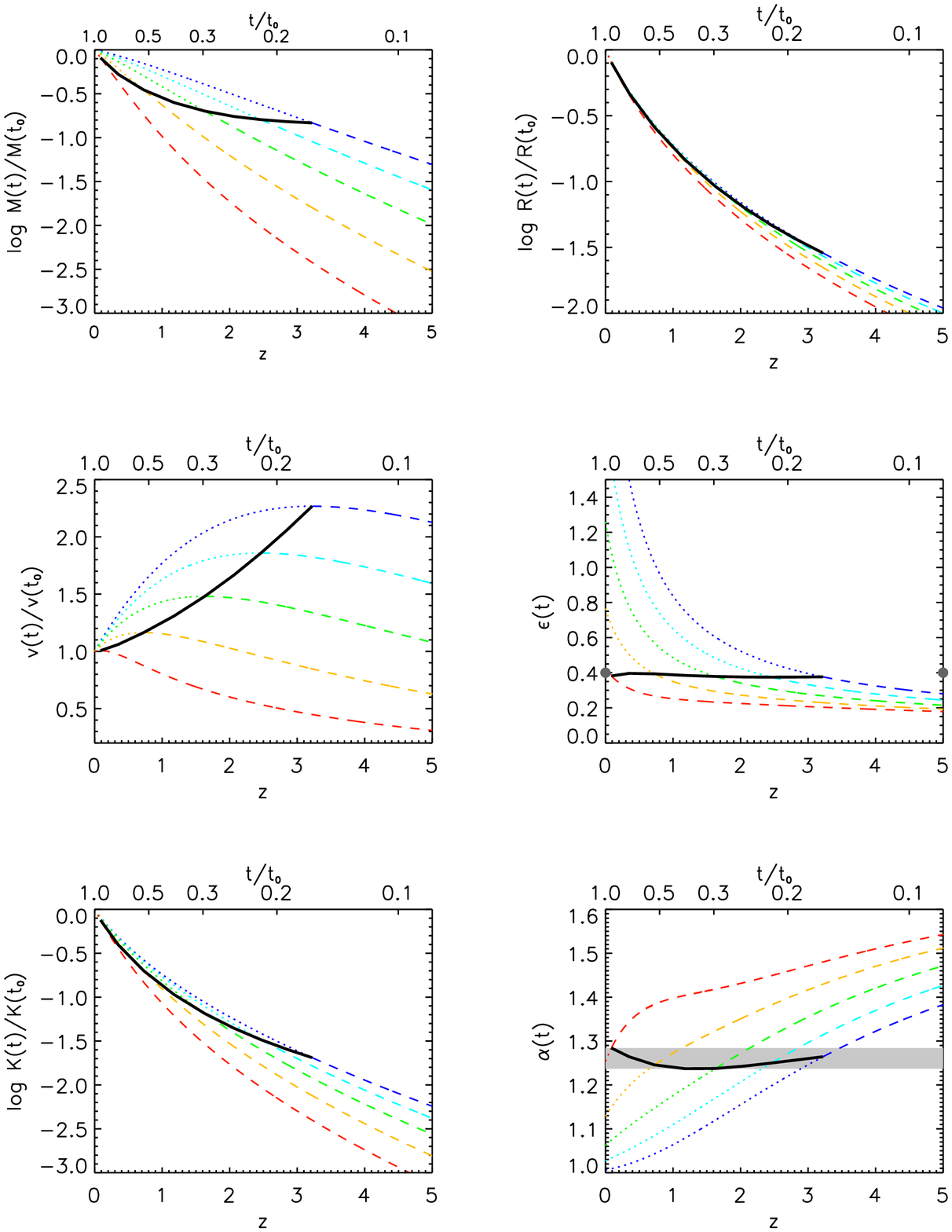}\caption{Evolution of the
various quantities along the average accretion history of DM
halos: mass $M$, radius $R$, velocity $v$, (inverse) mass
growth rate $\epsilon=-\mathrm{d}\log (1+z)/\mathrm{d}\log M$,
entropy $K$, and slope $\alpha\equiv\mathrm{d}\log
K/\mathrm{d}\log R$. The curves are for different current halo
masses in the range $10^{11}-10^{15}\, M_{\odot}$; from top to
bottom, blue lines refers to $10^{11}\, M_{\odot}$, cyan to
$10^{12}\, M_{\odot}$, green to $10^{13}\, M_{\odot}$, orange
to $10^{14}\, M_{\odot}$, and red to $10^{15}\, M_{\odot}$.
Lines are \textit{dashed} in the fast accretion regime, and
\textit{dotted} in the slow accretion phase; the \textit{solid}
line marks values at the transition epoch, see \S~3.3 for
details. In the bottom right panel, the \textit{shaded} area
highlights the narrow range expected for the values of $\alpha$
at the transition.}
\end{figure*}

\subsection{A heuristic approach}

Primarily we look for a link of the entropy slope $\alpha$ with
the mass growth rate in simple terms; this analysis will have
only heuristic scope, introductory to the full study performed
in \S~3.2. Toward our scope, we make contact with the classic
line of developments (Fillmore \& Goldreich 1984; Lu et al.
2006) that analytically relate the collapse histories to the
\textit{shape} and growth rate of the primordial DM
perturbations. The former may be expressed in terms of $\delta
M/M \propto M^{-\epsilon}$; but since a halo virializes when
$\delta M/M$ attains a critical threshold (e.g., $1.686$ in the
critical universe), the shape parameter
$\epsilon=-\mathrm{d}\log (1+z)/\mathrm{d}\log M$ effectively
governs also the (inverse) mass growth rate defined in terms of
the scaling $M\propto (1+z)^{-1/\epsilon}$.

Then we use the standard relations between redshift $z$ and
cosmic time $t$, that may be expressed in the form
$(1+z)\propto t^{-q}$. In the critical cosmology $q=2/3$ holds;
in the canonical Concordance Cosmology the value $q\approx 2/3$
still applies to $z \ga 1$, but as $z$ decreases $q$
progressively rises to values $q\approx 0.8$ for $z\la 0.5$.
Then the mass growth history writes simply $M\propto
t^{q/\epsilon}$ and the growth rate reads
$\dot{M}/M=q/\epsilon\,t$; from Eqs.~(8), simple algebra yields
the scaling laws $R\propto t^{(2+q/\epsilon)/3}$, $\rho\propto
t^{-2}$, and $K\propto t^{2\,(1+q/\epsilon)/3}$. On eliminating
$t$ one obtains $K\propto R^{\alpha}$ and $\rho\propto
R^{-\gamma}$ with slopes
\begin{equation}
\alpha=1+{1\over 1+2\,\epsilon/q}~~~~~\mathrm{and}~~~~~~\gamma={6\,\epsilon\over 2\,\epsilon+q}~,
\end{equation}
the latter being related to the former by the simple scaling
link $\gamma=6-3\,\alpha$, as expected at $r\approx r_0$ from
Eq.~(4). From Eq.~(9) $\alpha$ is seen to be slightly larger
than unity by a quantity slowly depending on $\epsilon/q$; this
is consistent with $K\propto R\, M^{1/3}\propto
R^{2-\gamma/3}$, the straightforward message from Eqs.~(8) with
$M\propto R^{3-\gamma}$.

These relationships for $\alpha$ and $\gamma$ embody in a
simple form some relevant cosmogonic and cosmological
information. For $z\ga 1$ where $q\approx 2/3$ applies, one
explicitly has $\alpha\approx (2+3\,\epsilon)/(1+3\,\epsilon)$
and relatedly $\gamma\approx 9\,\epsilon/(1+3\,\epsilon)$, the
latter being the well-known expression for the similarity
solutions to the spherical infall model in a critical Universe.
For $z\la 0.5$ instead when $q\approx 4/5$ applies, one has
$\alpha=(4+5\,\epsilon)/(2+5\,\epsilon)$ and $\gamma\approx
15\,\epsilon/(2+5\,\epsilon)$. As said, the remaining parameter
$\epsilon$ governs both the shape of the initial perturbations
and the times of their collapse. For instance, if the standard
top-hat shape applied with $\epsilon\approx 1$ corresponding to
a monolithic collapse, from the above relations one would
obtain $\alpha\approx 1.25$ and $\gamma\approx 2.25$ for halos
that virialize at $z\ga 1$, as pointed out early on by
Bertschinger (1985); for more massive halos that virialize at
$z\la 0.5$, from the scaling laws one anticipates a slightly
\emph{steeper} $\alpha\approx 1.29$ and a \emph{flatter}
$\gamma\approx 2.15$.

Qualitatively, different values of the slope $\gamma$ may be
related to other values of $\epsilon$, such as occurring in an
articulated shape of the initial perturbations, that also
yields a sequence of collapse times. With a realistic
bell-shaped perturbation $\epsilon\approx 1$ still applies to
its middle; in the outer regions $\epsilon\ga 1$ implies the
accretion to become slower, and correspondingly the density
slope $\gamma$ to steepen there relative to the middle one. The
central regions are described by $\epsilon \ll 1$,
corresponding to fast collapse; the scaling law then suggests
the density slope to be flatter there.

However, this simple approach cannot be carried very far. At
small radii also angular momentum effects come into play to
lower the density profile; they introduce another radial
scaling related to centrifugal barriers, as discussed in detail
by Lu et al. (2006); these authors show how a density slope
around $-1$ is enforced if efficient isotropization of
particles' orbits takes place during the fast collapse, while
still flatter slopes occur with more even angular momentum
distributions. These, however, must be consistent with the
overall equilibrium requirements; these orbital features are
actually subsumed into the macroscopic Jeans Eq.~(2) with
$K\propto r^\alpha$ and provide the running
$\gamma(r)=3\,\alpha/5+3\,v_c^2(r)/5\,\sigma^2(r)$ (the last
term is shown in Fig.~3) that is to replace the stiff scaling
value $\gamma=6-3\,\alpha$ above, but still recovers it at the
single radius $r_0$ (cf. Eqs.~[4] and [6]).

So we will \emph{depart} from the above heuristic approach, and
focus on computing the values of $\alpha$ to be inserted into
Jeans, also implementing precise mass and time dependencies of
the growth rate $\epsilon$. To focus these sensitive issues a
refined analysis and a careful discussion are required and
performed next.

\subsection{Advanced, semianalytic analysis}

We derive a robust time evolution and mass dependence of
$\dot{M}$ from an updated version of the Extended Press \&
Schechter formalism (EPS, see Lacey \& Cole 1993).

We are mainly interested in the average evolutionary behavior
of halos, so we compute the mean growth rate $\dot{M}$ of the
current mass $M$ as
\begin{equation}
\dot{M} =\int_0^{M}{dM'}~(M-M')\,{\mathrm{d}^2P_{M'\rightarrow
M}\over \mathrm{d}M'\mathrm{d}t}~,
\end{equation}
in terms of the differential \emph{merger rate}
$\mathrm{d}^2P_{M'\rightarrow M}/\mathrm{d}M'\mathrm{d}t$ given
in Appendix A. This incorporates the full cosmological
framework, and the detailed cold DM perturbation spectrum; as
to the former we implement the Concordance Cosmology, and as
for the latter we adopt the shape given by Bardeen et al.
(1986) corrected for baryons (Sugiyama 1995), and normalized as
to yield a mass variance $\sigma_8=0.8$ on a scale of
$8\,h^{-1}$ Mpc. We improve over previous studies of this kind
(e.g., Miller et al. 2006; Li et al. 2007; Salvador-Sol\'e et
al. 2007) by using the merger rate appropriate for ellipsoidal
collapses (Sheth \& Tormen 2002) as provided by Zhang et al.
(2008); this is shown by the latter authors to yield formation
times and growth rates in much closer agreement with numerical
simulations than the standard EPS theory. Further details are
presented in Appendix A.

Eq.~(10) provides the relation $\dot{M}(M,t)$ illustrated in
Fig.~5. This may be viewed as a first order differential
equation for $M(t)$, that is numerically solved once the mass
$M(t_0)$ at the present time $t_0$ has been assigned in the way
of a boundary condition. In terms of $M(t)$ so computed, the
evolution of the halo properties are obtained through Eqs.~(8);
the results are illustrated in Fig.~6, where we plot the time
and redshift evolutions for the inverse growth rate $\epsilon$,
the overall radius $R$, the mass $M$, the characteristic
velocities $v\propto \sigma$, the entropy $K$, and finally the
corresponding slope $\alpha\equiv\mathrm{d}\log
K/\mathrm{d}\log R$. The curves are for different
\emph{current} values of the halo masses $M(t_0)$ ranging from
$10^{11}$ to $10^{15}\, M_{\odot}$.

In fact, in these detailed mass accretion histories one can
recognize a \emph{transition} epoch when the circular velocity
$v_c^2$ of the halo attains its highest value along an
accretion history (see Li et al. 2007). The transition
separates two stages: an early stage of fast \emph{collapse},
when $\epsilon\la 0.4$ corresponds to a growth rate still high
on the running Hubble timescale, and nearly constant; and a
later phase of slower \emph{accretion} when $\epsilon$ takes
off from the value $0.4$, i.e., the growth rate is low and
decreasing. Small halos with $M\la 10^{12}\,M_{\odot}$ have
high transition redshifts, hence are now well advanced into
their slow accretion stage; on the other hand, very massive
ones with $M\ga 10^{15}\,M_{\odot}$ have low transition
redshifts, so today they have just completed their fast
accretion stage. A similar behavior has been identified in a
number of recent numerical simulations (Zhao et al. 2003;
Diemand et al. 2007), and has been recognized also in other
semianalytic computations based on the simpler, less precise
EPS theory (see Li et al. 2007; Neistein et al. 2006).

\subsection{Results}

From Fig.~6 we also see that the values of $\alpha$ at the
transition are narrowly \emph{constrained} to within $1.27\pm
0.02$, i.e., a $1.5\%$ range. We note that this range explains
why narrowly constrained phenomenological values of $\alpha$
are found within the halo \emph{bulk} from many and different
simulations since the year 2000. In particular, our values turn
out to agree with the best fitting exponent of $Q(r)$ [or
$K(r)$] with its scatter recently given in detail by Ascasibar
\& Gottl\"{o}ber (2008) within the halo bulk during calm
infall; these authors warn against extrapolating their results
to early times and outer radii.

Correspondingly, in our findings the values of $\alpha$ before
the transition should not be overinterpreted nor inserted into
the Jeans equation. In fact, at these early epochs the halos
experience rapid changes and are clearly out of equilibrium;
during this stage the $N-$body experiments (see Zhao et al.
2003; Peirani et al. 2006; Hoffman et al. 2007) highlight
turmoil associated with fast collapse and major mergers,
effective dynamical relaxation and mixing of DM particles.
Taking up from \S~3.1, it is during these events that the
orbits' angular momentum must set to a near isotropic
distribution such as to lower the inner densities.

But as the turmoil associated with the fast collapse subsides,
the orbital structure is bypassed and replaced by the
macroscopic entropy distribution $K\propto r^\alpha$ with
uniform values of $\alpha$, as phenomenologically found to hold
throughout the now stable halos' bulk. Then the values of
$\alpha$ at the transition epoch and the corresponding edge
$R=r_0$ can be derived from the scaling laws and applied to the
whole inner region, to obtain $\gamma(r)=
3\,\alpha/5+3\,v_c^2(r)/5\,\sigma^2(r)$, taking on the value
$6-3\, \alpha$ \emph{only} at the point $r_0$ according to the
results of \S~2.1.

The value of $\alpha$ is expected to persist in the bulk during
the subsequent development in the slow accretion stage, when
mass is added smoothly and mainly onto the outskirts, with
little mixing. Such a development is borne out by the widely
shared scenario of halo buildup from the inside out, championed
by Bertschinger (1985), Taylor \& Navarro (2001), and many
others up to Salvador-Sol\'e et al. (2007). Specifically, the
analysis by the former authors concerning the randomization and
the stratification of particles' orbits during secondary infall
indicates that their internal disposition is little affected by
subsequent, smoother mass additions on top of a closely static
potential well. Thus during the late, slow accretion stage the
structure for $r\la r_0$ will be unaffected.

Meanwhile, the \emph{outer} scale $R\propto M/\dot{M}^{2/3}$
beyond $r_0$ is increasingly stretched out to substantially
larger values (see Fig.~6, top right panel); correspondingly,
we find the outer slope $\alpha(R)\equiv \mathrm{d}\log
K/\mathrm{d}\log R$ at $R > r_0$ to decline in time (see
Fig.~6, bottom right panel). This translates into a radial
decrease $\alpha(r)$ on maintaining entropy stratification also
into these regions (see Bertschinger 1985, Taylor \& Navarro
2001). Such a decrease is to be expected also from considering
that $K(r)\propto r^{2-\gamma/3}\rightarrow r$ applies, on the
basis of the related asymptotic behaviors of $\rho(r)\propto
r^{-\gamma}$ and $\sigma^2(r)\propto r^{2-\gamma}$ where
$\gamma\rightarrow 3$. We find the values of $\alpha$ to
decrease by less than $2\%$ out to $r\approx 2\,r_0$, and less
than $10\%$ out to $r\approx 5\,r_0$.

Concerning the Jeans equation, this too (with its boundary
conditions set at $r=0$, see \S~2.2) technically works from the
inside out, and may be extended to the range $r > r_0$ with
decreasing slope $\alpha(r)$. The corresponding density run is
somewhat steepened for $r>r_0$, as if shifting from the upper
to the lower curves in Fig.~2; we have checked that the outer
density profile lowers by less than $20\%$ for $r \la 5\,r_0$.
State-of-the-art simulations still yield little information on
the outskirts for $r\ga r_0$ owing to low densities and noisy
data there discussed by Ascasibar \& Gottl\"{o}ber (2008). We
present elsewhere the complete solutions outside $r_0$ as a
testbed for future simulations aimed at resolving the
outskirts, and at fitting gravitational lensing observations
(see discussion in \S~5).

In a similar vein, we predict a minor increase of $\alpha$ at
the transition from $1.25$ to $1.29$ in moving from galaxy- to
cluster-sized current masses. A fit accurate to better than
$0.1\%$ to this average relation is provided by $\alpha\simeq
2.71-0.23\,\log(M/M_{\odot})+0.009\, [\log(M/M_{\odot})]^2$; in
particular for galaxy clusters, this implies values of
$\alpha\approx 1.27$ for Abell richness classes
$\mathfrak{r}\approx 0$ and $\alpha\approx 1.29$ for richness
class $\mathfrak{r}\ga 3$. The correlation $\alpha(M)$ may help
to understand the differences in the values of $\alpha$ found
in numerical simulations; in fact, the highest value $1.29$ to
date has been reported by Rasia et al. (2004) in looking at
cluster-sized halos, while the value $1.25$ first found by
Taylor \& Navarro (2001) refers to galaxy-sized halos.
Observational evidence supporting the above picture is
presented in \S~5.

To sum up, from the cosmological buildup we find for the
entropy slope the narrow allowed range
\begin{equation}
\alpha\approx 1.27\pm 0.02~;
\end{equation}
this we use in our Jeans-like description of DM equilibria
within the halos' \emph{bulk}. The bounds on $\alpha$ are
relevant in two respects: the upper one strengthens the formal
limit $35/27=1.\overline{296}$ required for the solutions of
the static Jeans equation to have viable profiles in the far
outskirts (see \S~2.2 and discussion in \S~5); the lower one
complementarily restricts the range of the viable values.

\section{A physical meaning for
\lowercase{$\kappa_{\mathrm{crit}}$}}

Here we propose to discuss the physical meaning of the quantity
$\kappa_{\mathrm{crit}}$ (defined in \S~2.1 following Taylor \&
Navarro 2001) in terms of a direct energetic argument. We first
note that given the asymptotics of the $\alpha$-profiles, the
velocity dispersion $v_c^2(r)$ must peak at a point $r_p\ga
r_0$ (see Fig.~4) to the value $v_p^2=4\pi G\rho(r_p)\,r_p^2$;
at this easily identified point the constant $\kappa$ also
writes $4\pi G\rho(r_p)\,r_p^2/\sigma_p^2=v_p^2/\sigma_p^2$,
and thus is seen to compare the estimate $\sigma_p^2$ for the
random kinetic energy with the estimate $v^2_p$ for the
gravitational potential. We then recall from \S~2.1 that a halo
in the viable equilibrium condition for a given $\alpha$ is
marked by a maximal value $\kappa_{\mathrm{crit}}(\alpha)$,
that is, by a \emph{minimal} level of randomized energy
$\sigma^2_p = v_p^2/\kappa_{\mathrm{crit}}$.

We now show that to a very good approximation the value for
$\kappa_{\mathrm{crit}}$ may be obtained from considering
\emph{incomplete} conversion of the infall kinetic energy
$v_{\mathrm{inf}}^2/2$ of DM particles into their dynamically
randomized energy $3\,\sigma^2_p/2$ during formation; a process
introduced by Bertschinger (1985), different from the complete
violent relaxation proposed by Lynden-Bell (1967) and discussed
widely, including the recent Arad \& Lynden-Bell (2005), Trenti
\& Bertin (2005) and Trenti et al. (2005).

We propose to derive the critical value of
$\kappa=v_p^2/\sigma_p^2$ from the relation
\begin{equation}
3\,\sigma_p^2\approx \eta\,
v_{\mathrm{inf}}^2~,~~~~~\mathrm{with}~~~~~\eta={3\over
2}\, {v_p^2/v_{\mathrm{inf}}^2\over
1+v_p^2/v_{\mathrm{inf}}^2}~.
\end{equation}
This expression for the effective conversion efficiency $\eta$
has been constructed on considering that: (i) it has to depend
on the ratio of the ($1-$D) infall velocity squared
$v_{\mathrm{inf}}^2$ to the depth of the potential well as
measured by the ($2-$D) circular velocity squared $v_p^2$; (ii)
the formal limit $v_p^2=2\, \sigma_p^2$ corresponding to
$\kappa=2$ (the isothermal powerlaw, see \S~2.1) must be
approached for a fast inflow with $v_{\mathrm{inf}}^2\gg
v_p^2$, which quantifies the incompleteness of the energy
conversion during the transit of a bullet with crossing time
short compared with the halo's dynamical time.

As a simple example guided by the canonical theory of
gravitational interactions (see Saslaw 1985; Cavaliere et al.
1992) we may consider that $v_{\mathrm{inf}}^2/v_p^2\approx 3$
is to hold for having considerable energy transfer; using this
simple approximation in Eq.~(12) yields
$\kappa_{\mathrm{crit}}\approx 2.\overline{6}$, a value
pleasingly close to $2.64$ as obtained numerically in \S~2.2
for the particular $\alpha$-profile with $\alpha= 1.25$. On the
other hand, this simple argument cannot pinpoint values of
$\kappa_{\mathrm{crit}}$ to the precision required to
discriminate different values of $\alpha$.

So, we may test our proposed $\eta$ as given in Eq.~(12) for
different values of $\alpha$ by using instead energy
conservation to express the infall velocity
$v_{\mathrm{inf}}^2\approx2\, \Delta \phi_{pt}$ in terms of the
adimensional potential drop $\Delta \phi_{pt}$ (normalized to
$v_p^2$) from the turnaround radius to $r_p$ (see Bertschinger
1985; Lapi et al. 2005). Inserting this into Eq.~(12) yields
the expression
\begin{equation}
\kappa_{\mathrm{crit}}\equiv {v_p^2 \over \sigma_p^2} =
2+{1\over \Delta \phi_{pt}}~;
\end{equation}
this highlights the deviation of $\kappa_{\mathrm{crit}}$ from
the value $2$ that would apply for the isothermal powerlaw (see
\S~3). Using the runs of $\phi$ appropriate for the
$\alpha$-profile as given in Fig.~2, we compute values for
$\kappa_{\mathrm{crit}}$ as a function of $\alpha$; note that
these decrease for increasing $\alpha$ which implies larger
$\Delta\phi_{pt}$, see Fig.~2. Fig.~7 shows how well Eq.~(13)
recovers the $\kappa_{\mathrm{crit}}(\alpha)$ relation obtained
through numerical integration of the Jeans equation.

This excellent fit to $\kappa_{\mathrm{crit}}$ means that the
expression for $\eta$ in Eq.~(12), more than a successful
guess, is describing what actually occurs in the macroscopic
equilibria expressed by the Jeans equation in keeping with the
numerical simulations. At the referee's suggestion we add that
what ultimately the first Eq.~(12) describes with the isotropic
form of its l.h.s. is a partial effective conversion at
$r\approx r_p$ from radial infall into tangential motions.

\begin{figure}
\epsscale{1.2}\plotone{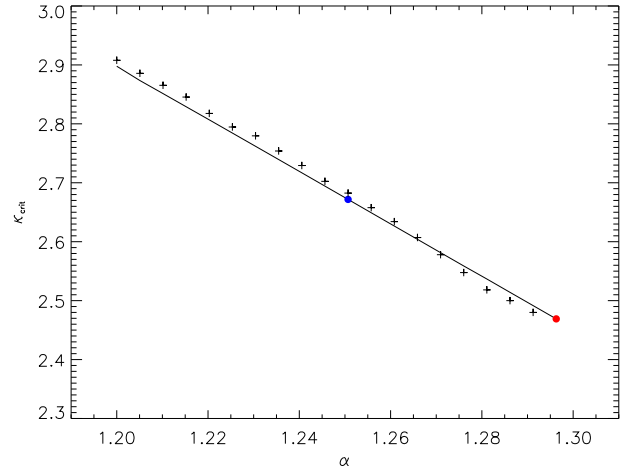}\caption{The
$\kappa_{\mathrm{crit}}(\alpha)$ relation, as in Fig.~1 (bottom
right panel). Crosses illustrate the goodness of our
approximation to $\kappa_{\mathrm{crit}}(\alpha)$ based on a
direct energetic argument, see \S~4 for details.}
\end{figure}

\begin{figure*}
\epsscale{0.8}\plotone{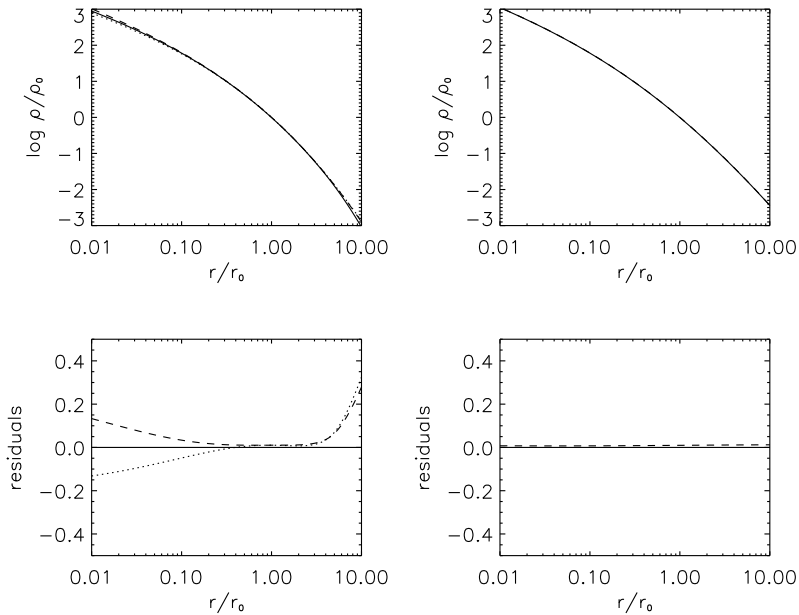}\caption{Analytic fits to
the density runs (top panels) and residuals (bottom panels) for
the $\alpha$-profiles with $\alpha=1.25$ (left panels) and
$\alpha=35/27$ (right panels). \textit{Solid} lines illustrates
the exact solutions from numerical integration of the Jeans
equation, \textit{dashed} lines show our analytic fitting
formula Eq.~(B1), and \textit{dotted} line (only in left
panels) is the fitting formula by Taylor \& Navarro (2001,
their Eq.~[10]); note that Eq.~(B1) for $\alpha=35/27$
coincides with the exact solution given by Dehnen \& McLaughlin
(2005).}
\end{figure*}

\section{Discussion and conclusions}

The general picture of DM halo formation as it emerges from
recent $N-$body simulations envisages a progressive buildup
through a two-stage development. At early epochs major mergers
associated with fast reshaping of the bulk gravitational
potential cause particles' orbits to dynamically relax and mix
into a definite equilibrium disposition. At later epochs the
halo outskirts progressively build up through rare if
persisting minor mergers and by smooth mass accretion, settling
to a quasi-static equilibrium.

The stable equilibria soon after the early stage are
effectively described in terms of a phenomenological entropy
run $K\propto r^\alpha$ with uniform slope $\alpha$ somewhat
larger than $1$. Spurred by this powerlaw behavior, one might
look for powerlaw solutions of the Jeans equation for the
equilibrium density profiles, only to find these just tangent
to the actual solutions. From here our analysis goes beyond
powerlaw densities, and provides two related blocks of novel
results.

First, we have computed and illustrated a wide set of novel
solutions to the isotropic Jeans equation, that we name
$\alpha$-\emph{profiles}, including as two particular instances
the prototypes investigated by Taylor \& Navarro (2001) and
Dehnen \& McLaughlin (2005). We have stressed that at given
$\alpha$ the solutions form a tangent bundle with $\rho(r)$
progressively curved down as the gravitational pull measured by
$\kappa=v_p^2/\sigma_p^2$ increases above the value
corresponding to the pure powerlaw profile
$\gamma=6-3\,\alpha$; from such bundles the $\alpha$-profiles
stand out for having a shape maximally curved to the point of
complying with sensible outer and inner behaviors, i.e., finite
total mass and no central hole. This occurs for a maximal value
$\kappa_{\mathrm{crit}}(\alpha)$ of the gravitational pull
$v_p^2$ relative to the random component $\sigma_p^2$, i.e.,
for a minimal $\sigma_p^2$ compared to $v_p^2$; this also
corresponds to the maximal slope $\gamma_p = 3\,\alpha/5 +
\kappa_{\rm crit}$ at the point $r_p$. Such solutions still
constitute a large set, being parameterized in terms of a wide
overall range $1< \alpha\leq 1.\overline{296}$ for the values
of $\alpha$.

Second, we have shown that toward further constraining these
values the key step is provided by the halos' cosmological
buildup; in fact, the scaling laws of \S~3 narrowly restrict
the viable \emph{range} to $\alpha \approx 1.25-1.29\approx
1.27\pm 0.02$, corresponding to current masses $M\sim
10^{11}-10^{15}\,M_{\odot}$. We have found these bounds to
apply at the epochs of \emph{transition} from fast collapse to
slow accretion, when the potential wells settle to their
maximal depths along an accretion history (see \S~3). In fact,
our semianalytic analysis not only recovers the two-stage
development observed in many recent $N-$body simulations, but
also explains \emph{why} only such narrowly constrained values
of $\alpha$ are actually found in many different numerical
works since the year 2000. The values $\kappa_{\mathrm{crit}}$
is correspondingly constrained to the range $2.7\ga
\kappa_{\mathrm{crit}}\ga 2.5$; we trace back these values to
limited \emph{randomization} for the infall kinetic energy of
DM particles during accretion, which leads to minimal values of
$\sigma^2$.

The $\alpha-$profiles from the Jeans equilibrium with $\alpha$
and $\kappa_{\mathrm{crit}}$ so constrained imply that galaxies
have a generally more \emph{compact} central structure than
groups and clusters, with appreciable \emph{deviation} from
self-similarity; on the other hand, the former have currently
developed extensive outskirts, while the very rich clusters do
not yet.

To embody in a handy expression the basic features of the
density runs for the $\alpha-$profiles, we provide here an
analytic fitting formula of the form often advocated on
phenomenological grounds (see Zhao 1996; Kravtsov et al. 1998;
Widrow 2000; Barnes et al. 2006), that reads
\begin{equation}
\rho(r)\propto {1\over \left({r/ r_0}\right)^{\gamma_a}\,
\left[1+w\,\left({r/ r_0}\right)^u\right]^{q}},
\end{equation}
with $\gamma_a$ governing the inner profile, $q$ the outer one,
$w$ and $u$ the position and sharpness of the radial
transition. Here for the first time such a profile is
\emph{substantiated} with parameters expressed in terms of the
derivatives of the Jeans equation (see \S~2.2) and of the
values for $\alpha$ as detailed in Appendix B. We stress that
the Jeans equlibrium (see Fig.~2) requires that the central
slopes are always \emph{flatter}, and the outer ones always
\emph{steeper} than in the empirical NFW fit. The goodness of
the above fitting formula is discussed in Appendix B, and
illustrated in Fig.~8.

\begin{figure}
\epsscale{1.2}\plotone{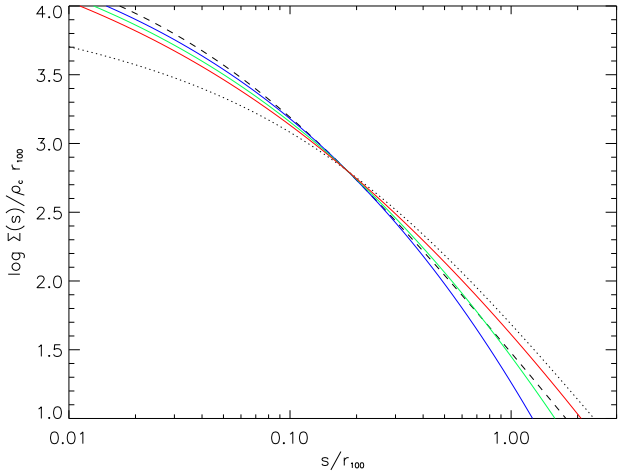}\caption{Surface density
runs from the $\alpha$-profiles for $\alpha=1.25$ (blue),
$\alpha=1.27$ (green) and $\alpha=1.29$ (red) and from the
empirical NFW profile with concentration parameter $c=5$ (black
dotted), compared with a fit to the gravitational lensing data
by Umetsu \& Broadhurst (2008, black dashed) in terms of an NFW
profile forced to high concentrations $c\approx 14$, but
actually overshooting the central data. Note how our physical
profiles do better both in the central and in the outer region
while relieving, along with the two-stage development, the
problems of high concentrations and short ages that arise from
NFW fits as discussed by Broadhurst et al. (2008).}
\end{figure}

We probe the physical relevance of the $\alpha-$profiles by
comparing in Fig.~9 the related projected density runs with the
results from recent, exquisite observations of strong and weak
gravitational lensing in and around massive galaxy clusters,
which just require \emph{flat} inner, and \emph{steep} outer
behaviors (Broadhurst et al. 2008, Umetsu \& Broadhurst 2008,
and references therein). It is seen how \emph{straightforward}
are our fits in terms of high-mass $\alpha-$profiles that
naturally exhibit just such features, compared to those in
terms of the NFW empirical formula that not only has problems
of ill-defined mass, but also would require particularly high
concentrations and old ages. Our findings substantiate the
$\alpha$-profiles, and stress their value as to specific
predictions for future simulations aimed at resolving, and for
weak gravitational lensing observations aimed at probing the
outer cluster densities. A full discussion of these issues
deserves a dedicated paper.

Two comments are in order concerning the validity of the Jeans
equation in the simple form of Eq.~(2). First, reasonable
anisotropies in the DM are described by the standard parameter
$\beta\equiv 1-\sigma_\theta^2/\sigma_r^2$ (Binney 1978), which
numerical simulations suggest to increase with $r$ from central
values $\beta(0)\approx -0.1 \div 0$ meaning near isotropy to
outer values close to $0.5$, meaning progressive outward
prevalence of radial motions (see Hansen \& Moore 2006). Such
anisotropies inserted into Jeans produce small variations of
the $\alpha-$profiles, to the effect of slightly increasing the
limiting value $\alpha = 35/27-4\, \beta(0)/27$; thus in the
related limiting solutions the inner slope is flattened
$\gamma_a = 3\,\alpha/5 + 6\,\beta(0)/5 =  7/9 +
10\,\beta(0)/9$, the middle slope $\gamma_0=6-3\,\alpha = 19/9
+ 4\,\beta(0)/9$ is steepened, and the outer slope $\gamma_b =
3\,(1+\alpha)/2 = 31/9-2\,\beta(0)/9$ is also steepened with
respect to the isotropic case, see Dehnen \& McLaughlin (2005).
The oppositely extreme case of total radial anisotropy with
constant $\beta=\beta(0)\approx 1$ discussed by Austin et al.
(2005) would yield a lower limiting $\alpha=31/27\approx
1.148$, while also producing a steep inner slope
$\gamma_a=17/9\approx 1.88$ (cf. \S~2.1). In brief, the
distinctive features of the $\alpha-$profiles are somewhat
enhanced, if anything, with radially increasing anisotropies.

Second, the static Jeans equation becomes increasingly
vulnerable in going into the outermost regions, since static or
even dynamic equilibrium is not granted beyond the virial
radius. Specifically, in the Concordance Cosmology the Hubble
expansion takes over at radii exceeding a few times the virial
radius where $\Omega_\Lambda\, (H_0\,r/v_c)^2\ga 1$ holds. This
weakens the case for the formal upper limit $\alpha= 35/27$ as
derived from Jeans analysis but correspondingly highlights the
relevance of the ceiling $\alpha\approx 1.3$ for a few
$10^{15}\, M_{\odot}$ halo we have derived from cosmogonic
evolution. On the other hand, any halo termination will be
uncertain by full factors of $2$; with flat outer density
slopes, as the case with $\alpha>35/27$ (see end of \S~2.1),
the related total masses will be ill-defined by similar
factors, corresponding for galaxy clusters to a disturbing
uncertainty by $\pm 2$ in Abell richness class (see Yee \&
L\'{o}pez-Cruz 1999).

Finally, we comment on our use of the DM `entropy'
$K(r)=\sigma^2(r)/\rho^{2/3}(r)$ -- actually the analog of the
thermodynamic adiabat, related to entropy by $s=3/2\, \log K$
-- rather than the equivalent functional `phase-space density'
$Q=K^{-3/2}$ -- that underlies the `information' $\log Q = - s$
with the obvious minus sign. Matter-of-factly, the pressure
gradient $\mathrm{d}\,(\rho\,\sigma^2)/\mathrm{d}r$ that
withstands the self-gravity in Eq.~(1) is most directly
expressed in terms of the macroscopic quantity $K$, in the form
$\mathrm{d}\,(\rho^{5/3}\,K)/\mathrm{d}r$. In the way of
interpretations, our lead is provided by the positive-definite
gradient $\Delta s = 3/2\,\Delta \log K$, and specifically by
its adimensional version $\mathrm{d}\log K/\mathrm{d}\log
r\equiv \alpha$.

We find $\Delta s$ to be convenient in understanding the
empirical evidence as to positive and uniform of $\alpha$
within the halos' bulk after the transition; then and there, in
fact, the particle orbits settle to a definite dynamical
regime, relaxed toward a periodic, ordered behavior
(Bertschinger 1985), corresponding to \emph{lower} entropy
layers at the center. In the outer regions, on the other hand,
the particle orbits stratify but mix poorly during the late,
slow buildup (see Bertschinger 1985; Taylor \& Navarro 2001)
with their apocenters distributed over larger and larger sizes;
we understand this dynamical regime in terms of $K$ or $\Delta
s$ spread over increasingly stretched radii $R>r_p$, resulting
in \emph{flatter} slopes.

It rests with the numerical values of $\alpha$ to translate the
fine orbital details into the macroscopic average of Jeans, as
clearly indicated by the consistency of the results from the
latter with semianalytic cosmogony and with simulations.

We end by stressing that the formulation of DM distribution in
terms of `entropy' -- despite the standard objections about
defining entropy for a collisionless medium out of thermal
equilibrium and dominated by long range weak gravity -- opens
an interesting perspective; this envisages direct comparisons
(see Faltenbacher et al. 2007) with the truly thermodynamic
entropy of the other major component in galaxy systems, namely,
the collisional plasma in local thermal equilibrium
constituting the intracluster medium. In fact, in the outer
regions these entropies though related as being both originated
by conversion of the infall energy under \emph{weak} gravity,
differ owing to their origin from \emph{different} conversion
processes (Lapi et al. 2005; Cavaliere \& Lapi 2006); in
addition, in the central region of most clusters the hot plasma
is heated up by the activity of supermassive black holes
originating under \emph{strong} gravity conditions.

\begin{acknowledgements}
Work supported in part by ASI and INAF. We thank Giuseppe
Bertin, Luigi Danese, Roberto Fusco-Femiano, Paolo Salucci and
Chiara Tonini for helpful discussions. We acknowledge the
constructive comments by an anonymous referee, especially
concerning the role of angular momentum and tangential motions
in \S~4 and 5.
\end{acknowledgements}

\begin{appendix}

\section{A. Merging rate of DM halo under ellipsoidal collapse}

To reliably compute the growth rate $\dot{M}$ in Eq.~(10), we
use the merging rate of DM halos under ellipsoidal collapse as
provided by Zhang et al. (2008); this reads
\begin{equation}
{\mathrm{d}^2P_{M'\rightarrow
M}\over \mathrm{d}M'\mathrm{d}t}={a_0\,
e^{-a_1^2\,\Delta\sigma^2/2\sigma_M^2}\over
\sqrt{2\pi}\,(\Delta\sigma^2)^{3/2}}\,
\left|{\mathrm{d}\sigma_{M'}^2\over
\mathrm{d}M'}\right|\,\left|{\mathrm{d}\delta_c\over
\mathrm{d}t}\right|\, \left\{1+a_2\,\left({\Delta\sigma^2\over
\sigma_M^2}\right)^{3/2}\, \left[1+{a_1\over \Gamma(3/2)}\,
\sqrt{\Delta\sigma^2\over \sigma_M^2}\right]\right\}
\end{equation}
where $\Delta\sigma^2=\sigma_{M'}^2-\sigma_{M}^2$ is the
difference of the mass variances corresponding to the masses
$M'$ and $M$, and
\begin{eqnarray}
&\nonumber & a_0\equiv 0.866\,\left\{1-0.133\,\left[\delta_c(z)^2/ \sigma^2_M\right]^{-0.615}\right\}~,\\
&\nonumber &\\
& & a_1\equiv 0.308\,\left[\delta_c(z)^2/ \sigma_M^2\right]^{-0.115}~,\\
&\nonumber &\\
&\nonumber & a_2\equiv 0.0373\,\left[\delta_c(z)^2/ \sigma^2_M\right]^{-0.115}~.
\end{eqnarray}
Note that the standard EPS expression for the spherical
collapse model (Lacey \& Cole 1993; Kitayama \& Suto 1996) is
recovered on setting $a_0=1$ and $a_1=a_2=0$.

Eqs.~(A1) and (A2) involve the critical threshold $\delta_c(z)$
for collapse, which is computed according to
\begin{equation}
\delta_c(z)=\delta_c\, {D(0)\over D(z)}
\end{equation}
in terms of the normalization value $\delta_c\approx
1.686\,[1+0.0123\,\log\Omega_M(z)]$ and of the growth factor
\begin{equation}
D(z)={5\over 2}\, {\Omega_M(z)\over 1+z}\, \left[{1\over 70}+{209\over 140}\,\Omega_M(z)-
{1\over 140}\,\Omega_M^2(z)+\Omega_M^{4/7}(z)\right]^{-1}~;
\end{equation}
in the above the quantity
$\Omega_M(z)=\Omega_M\,(1+z)^3/[1-\Omega_M+\Omega_M\,(1+z)^3]$
represents the evolved matter density parameter of the Universe
in the Concordance Cosmology.

Eqs.~(A1) and (A2) also involve the mass variance of the
primordial perturbation spectrum, that writes
\begin{equation}
\sigma_M^2 = 4\pi\,\int_0^{k_s}{\mathrm{d}k}\, k^2\, P(k)~,
\end{equation}
where $k_s=(6\pi^2\bar{\rho}_{M}/M)^{1/3}$ is the wavenumber
corresponding  to the mass $M$, and $\bar{\rho}_M\approx
2.8\times 10^{11}\,\Omega_M h^2\, M_{\odot}$ Mpc$^{-3}$ is the
mean background matter density of the Universe. The mass
variance is then normalized to the value $\sigma_8=0.8$ for the
mass corresponding to a scale of $8\,h^{-1}$ Mpc. As for the
power spectrum, we adopt the shape
\begin{equation}
P(k)\propto k\,T^2(k)
\end{equation}
in terms of the transfer function
\begin{equation}
T(k)\equiv {\log(1+2.34\,q)\over
2.34\,q}\,\left[1+3.89\,q+(16.1\,q)^2+(5.46\,q)^3+(6.71\,q)^4\right]^{-1/4}
\end{equation}
given by Bardeen et al. (1986). Here $q\equiv k/(h\,\Gamma)$
includes the shape correction factor $\Gamma=\Omega_M
h\,e^{-\Omega_b-\sqrt{2h}\,(\Omega_b/\Omega_M)}$ due to
baryons, after Sugiyama (1995); we recall that the values
$h=0.72$, $\Omega_M=0.27$, and $\Omega_b=0.044$ are adopted
throughout the paper.

\section{B. Fitting the density runs of the  $\alpha$-profiles}

Here we provide an explicit analytic fit to the
$\alpha$-profiles, whose numerical computation is sensitive to
precise initialization. We use the expression
\begin{equation}
\rho(r)\propto x^{-\gamma_a}\, (1+w\,x^u)^{-q}
\end{equation}
inspired by the exact solution for $\alpha=35/27$ given by
Dehnen \& McLaughlin (2005), and representing an extension of
the NFW (Navarro et al. 1997) formula; in the above $x\equiv
r/r_0$ is the adimensional radius, the inner slope has been
fixed to $\gamma_a=3\,\alpha/5$, while $u$, $q$, and $w$ are
parameters dependent on $\alpha$ to be determined.

To this purpose, we impose that at $r=r_0$ the density slope
from Eq.~(B1) and its first and second logarithmic derivatives
equal those of the $\alpha$-profiles. Thus we find
\begin{equation}
\gamma_a+u q\,{\,w\over w+1}=\gamma_0~~~~~~~~~~~~{u^2\,q\,w\over
(1+w)^2}=\gamma_0'~~~~~~~~~~~~{u^3\,q\,w\,(1-w)\over
(1+w)^3}=\gamma_0''~.
\end{equation}
Then we solve the above expressions for $w$, $u$, and $q$, to
obtain
\begin{equation}
w=1-{\gamma_0''\,(\gamma_0-\gamma_a)\over \gamma_0'^2}~~~~~~~~q={(\gamma_0-\gamma_a)^2\over
w\,\gamma_0'}~~~~~~~~u={(\gamma_0-\gamma_a)\,(w+1)\over q\, w}~.
\end{equation}

The values of $\gamma_0'$ and $\gamma_0''$ are easily computed
as a function of $\alpha$. First, Eqs.~(6) evaluated at $r=r_0$
give
\begin{eqnarray}
&\nonumber& \gamma_0'={3\over 5}\,\kappa_{\mathrm{crit}}(\alpha)+{2\over
3}\,(\gamma_0-\gamma_a)\, (\gamma_0-\gamma_b)\\
& &\\
&\nonumber&\gamma_0''={2\over
3}\,\gamma_0'\,(2\gamma_0-\gamma_a-\gamma_b)~;
\end{eqnarray}
then these quantities are easily recast in terms of $\alpha$ on
recalling from the main text (\S~2.2) that
$\gamma_a=3\,\alpha/5$, $\gamma_0=6-3\,\alpha$,
$\gamma_b=3\,(1+\alpha)/2$, and
$\kappa_{\mathrm{crit}}(\alpha)\simeq 8.228-4.443\,\alpha$.

To sum up, Eqs.~(B3) and (B4) provide the parameters
$w(\alpha)$, $u(\alpha)$, and $q(\alpha)$ as a function of
$\alpha$. E.g., for $\alpha=35/27$ we obtain $w=1.00$,
$u=0.44$, and $q=6.00$; thus our fitting formula recovers the
exact solution by Dehnen \& McLaughlin (2005). At the other
extreme, for $\alpha=1.25$ we obtain $w=0.23$, $u=0.39$, and
$q=20.77$; here our fitting formula approximates well (see
Fig.~8) the exact solution, improving upon the fit by Taylor \&
Navarro (2001, their Eq.~[10])\footnote{Note that the large
value of $q$ emulates the cutoff occurring for $\alpha=1.25$;
for $\alpha=35/27$ the lower $q=6$ emulates the asymptotic
falloff of this solution.}. In fact, we find that for
\emph{any} $\alpha$ between $1.25$ and $35/27$ our fit works to
better than a few $\%$ throughout the significant radial range
from $10^{-1}$ to several $r_0$; note that both the radial
range and the quality of the fit increase for increasing
$\alpha$, up to the point of pinning down the exact solution
for $\alpha=35/27$.

\end{appendix}


\begin{references}

\reference{}Arad, I., \& Lynden-Bell, D. 2005, MNRAS, 361, 385

\reference{}Ascasibar, Y. \& Gottl\"{o}ber, S. 2008, ApJ, 386,
2022

\reference{}Ascasibar, Y., Yepes, G., Gottl\"{o}ber, S., \&
M\"{u}ller, V. 2004, MNRAS, 352, 1109

\reference{}Austin, C.G., Williams, L.L.R., Barnes, E.I.,
Babul, A., \& Dalcanton, J.J. 2005, ApJ, 634, 756

\reference{}Bardeen, J., Bond, J., Kaiser, N., \& Szalay, A.
1986, ApJ, 304, 15

\reference{}Barnes, E.I., et al. 2007, ApJ, 654, 814

\reference{}Barnes, E.I., et al. 2006, ApJ, 643, 797

\reference{}Bertschinger, E. 1985, ApJS, 58, 39

\reference{}Binney J. \& Tremaine, S. 1987, Galactic Dynamics
(Princeton: Princeton Univ. Press)

\reference{}Binney J. 1978, MNRAS, 183, 779

\reference{}Broadhurst, T., et al. 2008, ApJ, 685, L9

\reference{}Cavaliere, A., \& Lapi, A. 2006, in Joint Evolution
of Black Holes and Galaxies, eds. M. Colpi, V. Gorini, F.
Haardt, and U. Moschella (Institute of Physics Publishing:
Bristol and Philadelphia).

\reference{}Cavaliere, A., Menci, N., \& Tozzi, P. 1999, MNRAS,
308, 599

\reference{}Cavaliere, A., Colafrancesco, S., \& Menci, N.
1992, ApJ, 392, 41

\reference{}Dehnen, W., \& McLaughlin, D.E. 2005, MNRAS, 363,
1057

\reference{}Diemand, J., Kuhlen, M., \& Madau, P. 2007, ApJ,
667, 859

\reference{}Faltenbacher, A., Hoffman, Y., Gottl\"{o}ber, S.,
\& Yepes, G. 2007, MNRAS, 376, 1327

\reference{}Fillmore, J.A., \& Goldreich, P. 1984, ApJ, 281, 1

\reference{}Hansen, S.H., \& Moore, B. 2006, NewA, 11, 333

\reference{}Hansen, S.H. 2004, MNRAS, 352, L41

\reference{}Hoffman, Y., Romano-D\'iaz, E., Shlosman, I., \&
Heller, C. 2007, ApJ, 671, 1108

\reference{}Kitayama, T., \& Suto, Y. 1996, MNRAS, 280, 638

\reference{}Kravtsov, A.V., Klypin, A.A., Bullock, J.S., \&
Primack, J.R. 1998, ApJ, 502, 48

\reference{}Lacey, C., \& Cole, S. 1993, MNRAS, 262, 627

\reference{}Lapi, A., Cavaliere, A., \& Menci, N. 2005, ApJ,
619, 60

\reference{}Li, Y., Mo, H.J., van den Bosch, F.C., \& Lin, W.P.
2007, MNRAS, 379, 689

\reference{}{\L}okas, E.L., \& Mamon, G.A. 2001, MNRAS, 321,
155

\reference{}Lu, Y., Mo, H.J., Katz, N., \& Weinberg, M.D. 2006,
MNRAS, 368, 1931

\reference{}Lynden-Bell, D. 1967, MNRAS, 136, 101

\reference{}Miller, L., Percival, W.J., Croom, S.M., \& Babic,
A. 2006, A\&A, 459, 43

\reference{}Navarro, J.F., Frenk, C.S., \& White, S.D.M. 1997,
ApJ, 490, 493

\reference{}Neistein, E., van den Bosch, F.C., \& Dekel, A.
2006, MNRAS, 372, 933

\reference{}Peebles, P.J.E. 1993, Principles of Physical
Cosmology, (Princeton, NJ: Princeton Univ. Press)

\reference{}Peirani, S., Durier, F., \& de Freitas Pacheco,
J.A. 2006, MNRAS, 367, 1011

\reference{}Rasia, E., Tormen, G., \& Moscardini, L. 2004,
MNRAS, 351, 237

\reference{}Salvador-Sol\'e, E., Manrique, A.,
Gonz\'alez-Casado, G., \& Hansen, S.H. 2007, ApJ, 666, 181

\reference{}Saslaw, W.C. 1985, Gravitational Physics of Stellar
and Galactic Systems (Cambridge: Cambridge Univ. Press)

\reference{}Sheth, R. K., \& Tormen, G. 2002, MNRAS, 329, 61

\reference{}Springel, V., Frenk, C.S., \& White, S.D.M. 2006,
Nature, 440, 1137

\reference{}Sugiyama, N. 1995, ApJS, 100, 281

\reference{}Taylor, J.E., \& Navarro, J.F. 2001, ApJ, 563, 483

\reference{}Trenti, M., \& Bertin, G. 2005, A\&A, 429, 161

\reference{}Trenti, M., Bertin, G., \& van Albada, T.S. 2005,
A\&A, 433, 57

\reference{}Umetsu, K., \& Broadhurst, T. 2008, ApJ, 684, 177

\reference{}White, S.D.M. 1986, in Inner Space/Outer Space: the
Interface between Cosmology and Particle Physics (Chicago:
Chicago Univ. Press), p. 228-245

\reference{}Widrow, L.M. 2000, ApJS, 131, 39

\reference{}Williams, L.L.R., Austin, C., Barnes, E., Babul,
A., \& Dalcanton, J.J. 2004, in Baryons in Dark Matter Halos,
eds. R. Dettmar, U. Klein, and P. Salucci (Trieste, Italy:
SISSA), see http://pos.sissa.it, p.20.1

\reference{}Yee, H.K.C., \& L\'{o}pez-Cruz, O. 1999, AJ, 117,
1985

\reference{}Zhang, J., Ma, C.-P., \& Fakhouri, O. 2008, MNRAS,
387, L13

\reference{}Zhao, D.H., Mo, H.J., Jing, Y.P., \& B\"{o}rner, G.
2003, MNRAS, 339, 12

\reference{}Zhao H. 1996, MNRAS, 278, 488

\end{references}
\end{document}